\documentclass[aps,pra,twocolumn,showpacs,superscriptaddress]{revtex4}
\usepackage{amsfonts}
\usepackage{amsmath}
\usepackage{amssymb}
\usepackage{epsfig}
\usepackage{graphicx}

\begin{document}

\title{Effective Quantum Dynamics of Interacting Systems with Inhomogeneous Coupling}

\date{\today}

\author{C. E. L\'{o}pez}
\affiliation{Departamento de F\'{\i}sica, Universidad de Santiago de Chile, Casilla 307
Correo 2, Santiago, Chile.}

\author{H. Christ}
\affiliation{Max-Planck-Institut f\"ur Quantenoptik, Hans-Kopfermann-Strasse 1, D-85748
Garching, Germany}

\author{J. C. Retamal}
\affiliation{Departamento de F\'{\i}sica, Universidad de Santiago de Chile, Casilla 307
Correo 2, Santiago, Chile.}

\author{E. Solano}
\affiliation{Max-Planck-Institut f\"ur Quantenoptik,
Hans-Kopfermann-Strasse 1, D-85748 Garching, Germany}
\affiliation{Physics Department, ASC, and CeNS,
Ludwig-Maximilians-Universit\"at, Theresienstrasse 37, 80333 Munich,
Germany} \affiliation{Secci\'on F\'{\i}sica, Departamento de
Ciencias, Pontificia Universidad Cat\'olica del Per\'u, Apartado
1761, Lima, Peru}

\pacs{42.50.Fx, 42.50.Vk, 73.21.La}

\begin{abstract}
We study the quantum dynamics of a single mode/particle interacting
inhomogeneously with a large number of particles and introduce an
effective approach to find the accessible Hilbert space where the
dynamics takes place. Two relevant examples are given: the
inhomogeneous Tavis-Cummings model (e.g., $N$ atomic qubits coupled
to a single cavity mode, or to a motional mode in trapped ions) and
the inhomogeneous coupling of an electron spin to $N$ nuclear spins
in a quantum dot.
\end{abstract}

\maketitle

\section{INTRODUCTION}

In quantum optics and condensed matter physics, a great effort has
been oriented to the study of single quantum systems, or a few of
them, and their experimental coherent control. This is the case of
cavity QED~\cite{reviewcavityQED}, trapped ions~\cite{reviewions},
quantum dots~\cite{Cerletti}, circuit QED~\cite{Blais}, among
others. On the other hand, many-particle physics describes a variety
of systems where collective effects may play a central role.

The interaction of a single atom with a quantized electromagnetic
mode, described by the Jaynes-Cummings (JC)
model~\cite{JaynesCummings}, has played a central role in quantum
optics and other related physical systems. Among its many
fundamental predictions, we can mention vacuum-field Rabi
Oscillations, collapses and revivals of atomic populations, and
squeezing of the radiation field~\cite{review1}. Atomic cloud
physics can be conveniently described by the Dicke
model~\cite{dicke1}, when considering the interaction of atoms with
light in free space, or by the Tavis-Cummings (TC)
model~\cite{tavis1}, when the coupling takes place inside a cavity.
In most treatments and applications of the TC model, a constant
coupling between the atoms and the radiation field is assumed,
simplification that greatly reduces the complexity of an analytical
description~\cite{cooperativeeffects,jc1,jc2,Retzker}. Even in the
case of many atoms, the homogeneous-coupling case profits from the
$SU(2)$ group structure of the collective operators, allowing at
least numerical solutions. However, the situation is drastically
different when we consider the more realistic case of inhomogeneous
coupling. In this case, there is no such straightforward way to
access the Hilbert space, because all angular momenta
representations are mixed through the dynamical evolution of the
system and no analytical approach is known.

On the other hand, electron and nuclear spin dynamics in
semiconductor nanostructures has become of central interest in
last years~\cite{Cerletti,Wolf}, and several techniques for
quantum information processing have been
proposed~\cite{Loss1,Taylor2,Privman,Kane,Levy,Ladd}. For example,
ensembles of polarized nuclear spins in quantum dots~\cite{Taylor}
and quantum Hall semiconductor nanostructures~\cite {Mani} have
been proposed as a long-lived quantum memory for the electron
spin. Recently, polarization procedures for the nuclear spins in a
quantum dot have been studied
theoretically~\cite{zoller,Christ06,DengHu}, and first experiments
have achieved considerable dynamical nuclear
polarization~\cite{Lei}. However, difficulties in producing a
fully polarized nuclear state suggest the necessity of analyzing
the effect of imperfect polarization. This treatment faces the
same theoretical challenges as the study of the inhomogeneous TC
model, because of the unavoidable inhomogeneous coupling of
electron and nuclear spins in quantum dots.

In this work, we develop an effective approach to study the dynamics
of a single mode/particle coupled inhomogeneously to a large number
of particles. In particular, we analyze in Sec.~\ref{ITC} the
general interaction of a collection of two-level atoms with a
quantized field mode, that is, the inhomogeneous TC model. We
identify a subset of the Hilbert space that is relevant for the
dynamics and show that we are able to reproduce accurate results
with a small number of states. In Sec.~\ref{sect_QD} we consider the
hyperfine interaction of a single excess electron spin confined in a
quantum dot with quasi-polarized nuclear spins. Using the formalism
developed in section~\ref{ITC}, we study the transfer of quantum
information between those systems in the presence of a single
excitation in the nuclear spin system prior to the write-in
procedure. In Sec.~IV, we present our concluding remarks.

\section{INHOMOGENEOUS TAVIS-CUMMINGS MODEL\label{ITC}}

The Hamiltonian describing the inhomogeneous coupling of $N$ atoms
with a quantized single field mode, in the interaction picture and
under resonant conditions, can be written as
\begin{equation}
\hat{H}_{\mathrm{ITC}}=\sum_{i=1}^{N}g_{i}\left( \hat{\sigma}
_{i}^{-} \hat{a}^{\dagger }+ \hat{\sigma}_{i}^{+} \hat{a}\right) .
\label{ham1}
\end{equation}
Here, $g_{i}=g({\vec{r}}_{i})$ is the (real) inhomogeneous coupling
strength of atom $i$ at position ${\vec{r}}_{i}$,
$\hat{\sigma}_{i}^{-}$ ($\hat{\sigma}_{i}^{+}$) is the lowering
(raising) operator of atom $i$, and $\hat{a}$ ($\hat{a}^{\dagger}$)
is the annihilation (creation) operator of the field mode. We will
refer to this model as the inhomogeneous Tavis-Cummings (ITC) model.
In the homogeneous case, $g_{i}=g$, $\forall i$, one can define as
angular momentum operator $ \hat{J}^{+}\propto g\sum_{i=1}^{N}
\hat{\sigma}_{i}^{+}$, describing transitions between the common
eigenvectors of $\hat{J}_{z}$ and $\hat{J}^{2}$, where
$\hat{J}_{z}=\sum_{i=1}^{N} \hat{\sigma}_{z_{i}}$. This is not true
for the inhomogeneous case, because
${\hat{\tilde{J}}^{+}}=\sum_{i=1}^{N}g_{i} \hat{\sigma}_{i}^{+}$
does not satisfy anymore the $SU(2)$ algebra. However, we will show
that this problem can still be addressed with certain restrictions.
The strategy consists in following the Hilbert space that the system
will visit along its evolution, and implementing a truncation that
will depend on the window evolution time.

Let us first consider the initial condition given by
\begin{equation}
|\Psi (0)\rangle =\sum_{n=0}c_{n}|n\rangle |\mathbf{\bar{0}\rangle },
\end{equation}
where $|n\rangle $ denotes a photonic Fock state and
$|\mathbf{\bar{0}\rangle }$ the collection of $N$ atoms in the
ground state. For each product state $ |n\rangle
|\mathbf{\bar{0}\rangle }$, we have a fixed number of excitations,
which, even in the inhomogeneous case, is conserved by the
dynamics determined by $\hat{H}_{\mathrm{ITC}}$. If the initial
state is $|0\rangle |\mathbf{ \bar{0}\rangle }$, the unitary
evolution is trivial
\begin{equation}
|0\rangle |\mathbf{\bar{0}\rangle \longrightarrow }|0\rangle
|\mathbf{\bar{0} \rangle }.
\end{equation}
Starting from initial state $|1\rangle |\mathbf{\bar{0}\rangle }$,
$\hat{H}_{ \mathrm{ITC}}$ produces a nontrivial dynamics via the
term $\hat{\tilde{J}}^{+} \hat{a}$, so that
\begin{equation}
\hat{\tilde{J}}^{+} \hat{a} |1\rangle |\mathbf{\bar{0}\rangle
=}|0\rangle
\sum_{i=1}^{N}g_{i}|\mathbf{\bar{1}}_{i}\mathbf{\rangle },
\label{eq1}
\end{equation}
where
$|\mathbf{\bar{1}}_{i}\mathbf{\rangle}=\hat{\sigma}_i^+|\mathbf{\bar{0}}\rangle$
represents a $N$-atom state with one excitation in the $i$th atom.
We define
\begin{equation}
|\mathbf{\bar{1}\rangle =}\frac{1}{\sqrt{\sum_{i=1}^{N}g_{i}^{2}}}
\sum_{i=1}^{N}g_{i}|\mathbf{\bar{1}}_{i}\mathbf{\rangle },
\label{one}
\end{equation}
a normalized state such that
\begin{equation}
\langle 0|\langle \mathbf{\bar{1}}|H_{\rm ITC}|1\rangle
|\mathbf{\bar{0}\rangle =} \sqrt{\sum_{i=1}^{N}g_{i}^{2}}\equiv
N_{0}.
\end{equation}
Note that the system evolves in a closed two-dimensional subspace
$\{|1\rangle |\mathbf{\bar{0}\rangle ,}$ $|0\rangle
|\mathbf{\bar{1}\rangle }\}$ in which all states have the same
fixed number of excitations.

\begin{figure}
\includegraphics[width=90mm]{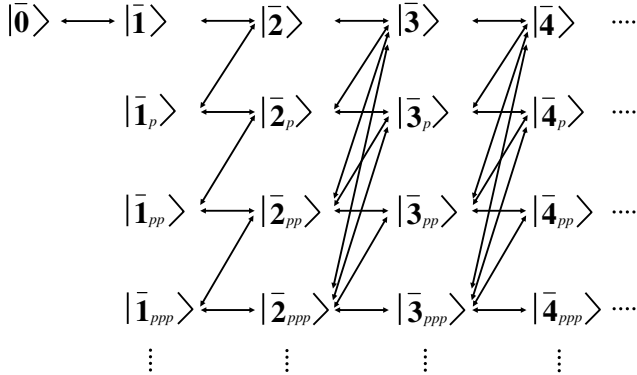}\vspace*{-0.8cm}
\caption{Schematic representation of the growth of the Hilbert
space associated with the collective atomic states.}
\label{couplings}
\end{figure}

If we start the dynamics with the 2-excitation state $|2\rangle
|\mathbf{\bar{0}\rangle }$, it evolves through
\begin{eqnarray}
\hat{\tilde{J}}^{+} \hat{a} |2\rangle |\mathbf{\bar{0}\rangle }
&\mathbf{=}&\sqrt{2}
|1\rangle \sum_{i=0}g_{i}|\mathbf{\bar{1}}_{i}\mathbf{\rangle }  \notag \\
&=&\sqrt{2}N_{0}|1\rangle |\mathbf{\bar{1}\rangle } .
\end{eqnarray}
The generated state $|1\rangle |\mathbf{\bar{1}\rangle }$ produces now
\begin{eqnarray}
\hat{\tilde{J}}^{+} \hat{a} |1\rangle |\mathbf{\bar{1}\rangle }
&\mathbf{=}&\frac{1}{N_{0}} |0\rangle
\sum_{i=1}^{N}g_{i}\hat{\sigma} _{i}^{+}\sum_{j=1}^{N}g_{j}|
\mathbf{\bar{1}}_{j}\mathbf{\rangle }  \notag \\
&=&\frac{N_{1}}{N_{0}}|0\rangle |\mathbf{\bar{2}\rangle } ,
\end{eqnarray}
where
\begin{equation}
|\mathbf{\bar{2}\rangle }\mathbf{=}\frac{2}{N_{1}} \sum_{i <
j}g_{i}g_{j}| \mathbf{\bar{2}}_{ij}\mathbf{\rangle } ,
\end{equation}
and
\begin{equation}
N_{1}=\sqrt{4\sum_{i < j}g_{i}^{2}g_{j}^{2}} .
\end{equation}
At the same time
\begin{equation}
\hat{\tilde{J}}^{-} \hat{a}^{\dagger } |1\rangle
|\mathbf{\bar{1}\rangle }\mathbf{=}\sqrt{2} N_{0}|2\rangle
||\mathbf{\bar{0}}\rangle .
\end{equation}
On the other hand, the state $|0\rangle |\mathbf{\bar{2}\rangle }$ evolves
according to
\begin{eqnarray}
\hat{\tilde{J}}^{-} \hat{a}^{\dagger } |0\rangle
|\mathbf{\bar{2}\rangle } &\mathbf{=}&|1\rangle \frac{ 2}{N_{1}}
\sum_{k}\sum_{i < j}g_{k}g_{i}g_{j}
\hat{\sigma}^{-}_{k}|\mathbf{\bar{2}
}_{ij} \mathbf{\rangle }  \notag \\
&=&|1\rangle |\phi _{1}\rangle ,
\end{eqnarray}
with
\begin{equation}
|\phi _{1}\rangle =\sum_{i}(\sum_{j\neq i}c_{ij}g_{j})|\mathbf{\bar{1}}_{i}
\mathbf{\rangle } ,
\end{equation}
and $c_{ij}=2g_{i}g_{j}/N_{1}$. The result is a one excitation
state in the atomic system which is different from
$|\mathbf{\bar{1}\rangle }$ defined in Eq.~(\ref{one}). The
resulting state can be expressed \ as a linear combination of
$|\mathbf{\bar{1}\rangle }$ and an orthogonal state
$|\mathbf{\bar{1}}_{p} \rangle$ that can be conveniently expressed
as
\begin{equation}\label{xxx}
|\mathbf{\bar{1}}_{p} \mathbf{\rangle =} \frac{1}{\sqrt{\langle
\phi _{1}|\phi _{1}\rangle -|\langle \mathbf{\bar{1}}|\phi
_{1}\mathbf{\rangle |}^{2}}} \mathbf{(}|\phi _{1}\rangle -\langle
\mathbf{\bar{1}}|\phi _{1}\mathbf{\ \rangle
}|\mathbf{\bar{1}\rangle )} ,
\end{equation}
yielding
\begin{equation}
|1\rangle |\phi _{1}\rangle =\langle \mathbf{\bar{1}}|\phi
_{1}\mathbf{\ \rangle }|1\rangle |\mathbf{\bar{1}\rangle
+}\sqrt{\langle \phi _{1}|\phi _{1}\rangle -|\langle
\mathbf{\bar{1}}|\phi _{1}\mathbf{\rangle |} ^{2} }|1\rangle
|\mathbf{\bar{1}}_{p}\mathbf{\rangle } .
\end{equation}
The state $|1\rangle |\mathbf{\bar{1}}_{p}\mathbf{\rangle }$ is a
new orthogonal state of the considered $2$-excitation Hilbert
subspace. Via $\hat{\tilde{J}}^{-} \hat{a}^{\dagger }$, $|1\rangle
|\mathbf{\bar{1}}_{p}\mathbf{\rangle }$ goes back to $|2\rangle |
\mathbf{\bar{0}\rangle }$, the latter being coupled only to the
state $|1\rangle |\mathbf{\bar{1}\rangle }$. We observe also that,
while $\hat{\tilde{J}}^{-} \hat{a}^{\dagger } |1\rangle
|\mathbf{\bar{1}}_{p}\mathbf{ \rangle =0}$, the associated Hilbert
subspace grows according to
\begin{equation}
\hat{\tilde{J}}^{+} \hat{a} |1\rangle
|\mathbf{\bar{1}}_{p}\mathbf{\rangle =\alpha }_{2}
\mathbf{|}0\rangle |\mathbf{\bar{2}\rangle +\alpha
}_{2p}\mathbf{|}0\rangle | \mathbf{\bar{2}}_{p}\mathbf{\rangle } ,
\end{equation}
where we introduced $|\mathbf{\bar{2}}_{p}\mathbf{\rangle }$
perpendicular to $|\mathbf{\bar{2}\rangle }$. The state
$|\mathbf{\bar{2}} _{p}\mathbf{ \rangle }$ thus obtained is coupled
via $\hat{\tilde{J}}^{-}$ to states $| \mathbf{\bar{1}}
_{p}\mathbf{\rangle }$ and $|\mathbf{\bar{1}}_{pp}\mathbf{\ \rangle
}$, where $|\mathbf{\bar{1}}_{pp}\mathbf{\rangle }$ is orthogonal to
$|\mathbf{ \bar{1}}_{p}\mathbf{\rangle }$, and so on. In fact,
applying $(\tilde{J}^{\dagger }a)^{n}$ on $\mathbf{|}n\rangle
|\mathbf{\bar{0} \rangle }$ yields the state
\begin{equation}
\mathbf{|}0\rangle |\mathbf{\bar{n}\rangle = } (n!)^{3/2} \, | 0
\rangle \!\!\!\!\! \sum_{i_{1} < ... < i_{n}} \!\!\!\!\! g_{i_{1}}
g_{i_{2}} ... g_{i_{n}} |\mathbf{\bar{n}}_{i_{1}i_{2}...i_{n}}
\mathbf{\rangle } , \label{Jnan}
\end{equation}
and the collective atomic states couple following the table of Fig.~\ref{couplings}.
In general, the states $\mathbf{|}n\rangle |\mathbf{\bar{m}\rangle } $ lead to linearly independent subspaces of conserved $n+m$ excitations.

\begin{figure}
\hspace*{-0.7cm}
\includegraphics[width=78mm,height=65mm]{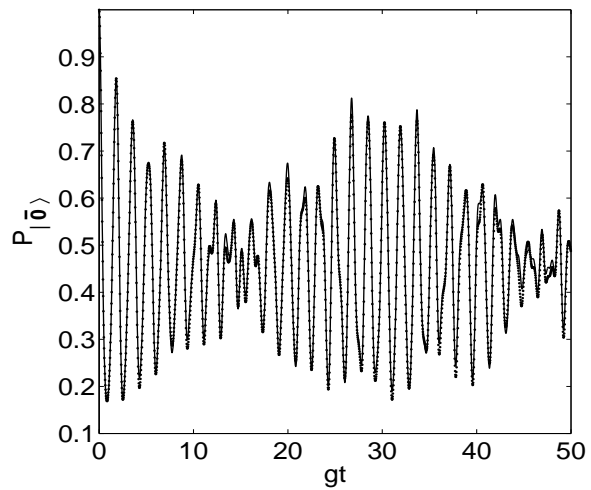}\vspace*{0.0cm}
\caption{Evolution of the collective atomic ground state population, $P_{| \mathbf{ \bar{0}} \rangle}$, in the
case of inhomogeneous coupling $g_{j}=g\sin \left[ j\protect\pi
/\left( N+1\right) \right] $, $j = 1, .., N$, with $N=6$, an
initial atomic state $\mathbf{\left\vert \mathbf{ \bar{0}}
\right\rangle }$, and an initial coherent state in the field with
$ \bar{n} =1.8$. Exact evolution (solid line) and approximated
evolution (dotted line).} \label{pob}
\end{figure}

\begin{figure}[h]
\hspace*{-0.3cm}
\includegraphics[width=85mm,height=78mm]{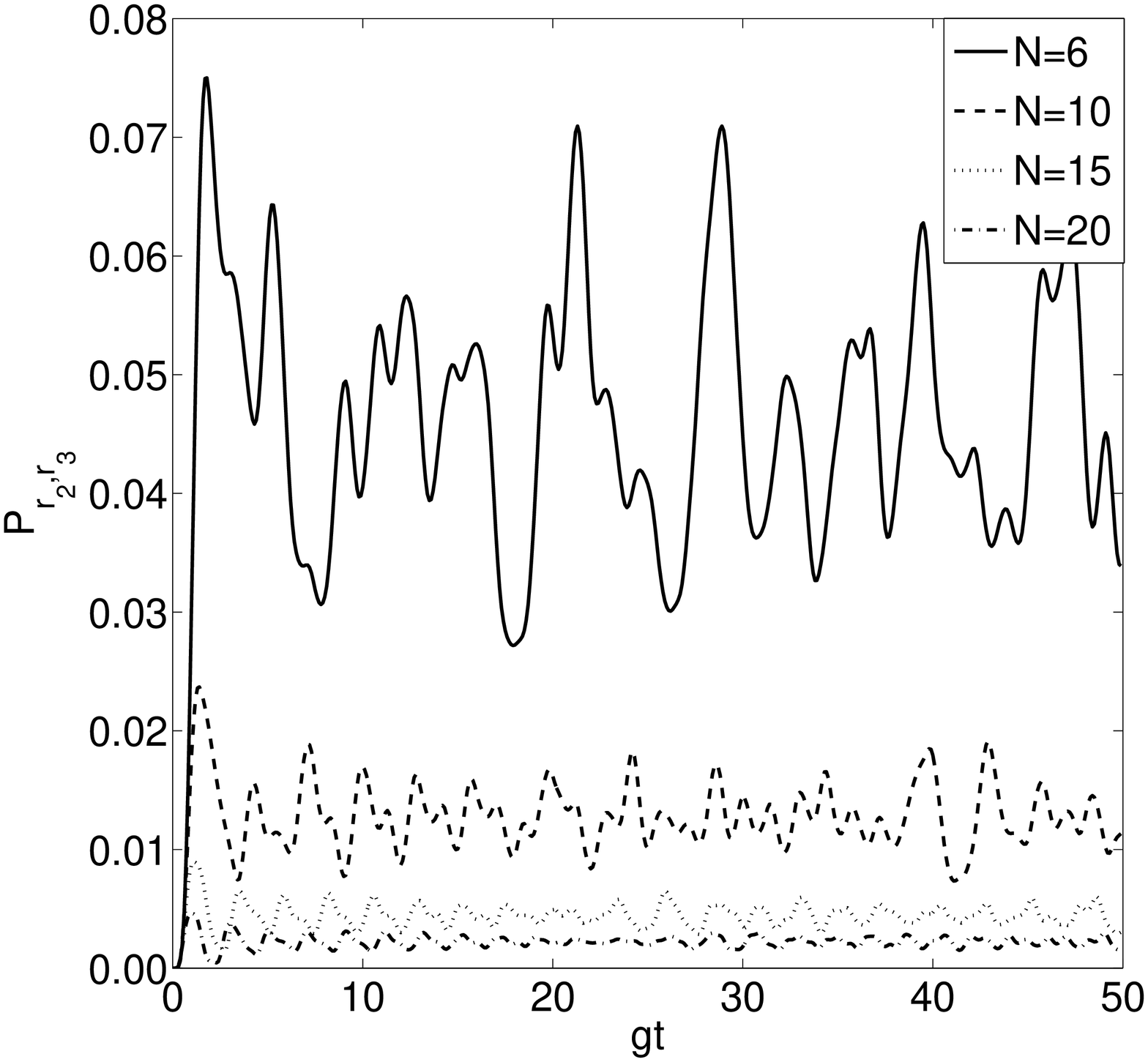}\vspace*{0.0cm}
\caption{Evolution of the total population of the second and third
rows, $P_{r_2,r_3}$, for different number $N$ of atoms.} \label{popN}
\end{figure}

Eventually, all states of the Hilbert space will be visited by an
arbitrary system evolution but, within a given time window, only a
restricted portion of the Hilbert space will be accessed. However,
in the example of Fig.~\ref{pob}, we observe that our
approximation for the evolution of the collective atomic ground
state is in good agreement with the exact result for very long
times, reproducing even  population collapses and revivals.
Surprisingly, to achieve this level of accuracy, it is enough to
consider the first two rows displayed in Fig.~\ref{couplings}. The
number of columns that has to be considered is determined by the
number of initial excitations in the system because the
Hamiltonian of Eq.~(\ref{ham1}) preserves this number. In
consequence, the low mean photon number, $\bar{n} = 1.8$, suggests
that the initially unpopulated atomic state may reach at most a
similar mean value. Given that the number of atoms is larger, $N =
6$, it is expected that the normalized couplings between the few
visited atomic states in the first row of Fig.~\ref{couplings} and
the rows below are quite small. In order to confirm this
intuition, we show in Fig.~\ref{popN} how the population of the
second and third rows of Fig.~\ref{couplings} decreases as the
number of atoms $N$ increases for the same parameters of
Fig.~\ref{pob}. Clearly, also increasing inhomogeneity of the
coupling causes increased leakage into higher order rows, see
Eq.~(\ref{xxx}). The example of Fig.~\ref{pob} shows a reduction of
the Hilbert space dimension from $2^6 \times 7 = 448$, where $7$
is the dimension of the truncated field space, to an effective
size of $49$. In general, to improve the accuracy, we only need to
take into account more rows in the calculation.

\begin{figure}
\vspace*{0.0cm} \hspace*{-0.6cm}\includegraphics[width=85mm,height=70mm]{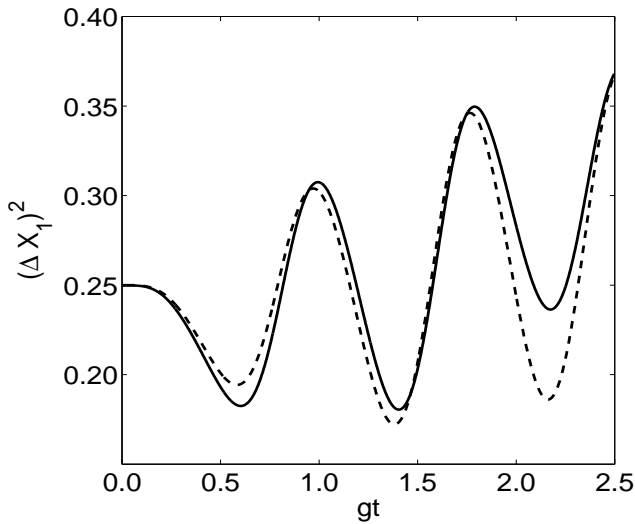} 
\caption{Evolution of the quadrature fluctuations $\left( \Delta
X_{1}\right) ^{2}=\left\langle X_{1}^{2}\right\rangle
-\left\langle X_{1}\right\rangle ^{2}$ for $N=8$, initial atomic
ground state, and a coherent field with $\bar{n}=1$. Solid line:
inhomogeneous coupling $g_{j}=g\sin \left[ j\pi /\left( N+1\right)
\right]$, where $j$ labels the position of the $j$-th atom. Dashed
line: homogeneous case with $g_j=1/(\sqrt{N})(\sum_{j}
g_j^2)^{1/2}$.} \label{sq8}
\end{figure}

We have proposed a method to follow the dynamics of a system with
inhomogeneous coupling in a relatively simple manner. It is now
interesting to see how strong can be the effects of inhomogeneity
on the dynamics of relevant observables. For example, in
Fig.~\ref{sq8}, we compare the different predictions for the
fluctuations of the field quadratures. For the homogeneous case,
the first time scale corresponds to $\sqrt N g$, while for the
inhomogeneous case the analogous would be $(\sum_j g_j^2)^{1/2}$.
In this manner, to compare both cases, the homogeneous and the
inhomogeneous we consider the homogeneous coupling case with
coupling $g_j=1/(\sqrt{N})(\sum_{j} g_j^2)^{1/2}$. In
Fig.~\ref{sq8} we observe that not only does the inhomogeneous
situation differ from the homogeneous in the typical timescale of
the dynamics, but also it shows additional effects, which, in the
case we consider, are reflected in an increase of the quadrature
fluctuations. Our method allows the study of these effects for
particle numbers that are non tractable for standard simulation
methods.

\section{Quantum Dots\label{sect_QD}}

Another interesting physical system involving inhomogeneous
coupling in a natural way is that of a single electron confined in
a quantum dot surrounded by nuclear spins. In presence of a
magnetic field $B_0$ along the $z$-axis the associated
Hamiltonian, neglecting dipolar interactions between nuclear
spins, can be written as~\cite{CoishLoss,Schliemann}
\begin{equation}
\hat{H}_{\mathrm{QD}}=g^{\ast }\mu _{B}B_{0}\hat{S}_{z}+g_{n}\mu
_{n}B_{0}\sum\limits_{i=1}^{N}\hat{I}_{z}^{(i)}+\hat{V}_{\mathrm{HF}}.
\label{hamil1}
\end{equation}
Here, the first two terms are the Zeeman energies of the electron
and nuclear spins, and $\hat{V}_{\mathrm{HF}}=\hat{V}_{D}+\hat{V}
_{\Omega }$ describes the hyperfine contact interaction of a single
electron spin interacting with $N$ nuclear spins, where $\hat{V}
_{D}=\sum_{i=1}^{N}\alpha _{i}\hat{I}_{z}^{(i)}\hat{S}_{z}$, and
$\hat{V} _{\Omega }=\sum_{i=1}^{N} (\alpha _{i}/2) \left(
\hat{I}_{-}^{(i)}\hat{S}_{+}+\hat{ I}_{+}^{(i)} \hat{S}_{-}\right)$.
The coefficients $\alpha _{i}=Av_{0}\left\vert \psi \left(
\vec{r}_{i}\right) \right\vert ^{2}$ are the inhomogeneous coupling
constants, where $A$ is the hyperfine coupling constant, $v_{0}$ is
the inverse density of nuclei in the material, and $\psi \left(
\vec{r}_{i} \right)$ is the envelope function of the electron at
position $\vec{r }_{i}$. The term $\hat{V}_{D}$ induces an effective
magnetic field for the electron $\hat{B}_{\mathrm{eff}}=B_{0} +
(1/g^{\ast }\mu _{B}) \sum_{i=1}^{N}\alpha _{i} \hat{I}_{z}^{(i)}$,
the so called Overhauser shift. Below, the nuclear Zeeman energies
will be neglected, as their magnetic moment is much smaller than the
one for the electron~\cite{Schliemann}.

By applying an appropriate field $B_{0}$, so that $g^{\ast }\mu
_{B}\hat{B}_{\mathrm{ eff}} \ll \hat{V}_{\Omega }$, one may allow
the spin-exchange term to dominate the dynamics~\cite{Taylor}. In
this case, and defining $\hat{A}_{\pm ,z}=\sum_{j}\alpha
_{j}\hat{I}_{\pm ,z}^{(j)}$, the Hamiltonian of Eq.~(\ref{hamil1})
approximately reduces to $\hat{V}_{\Omega} = (1/2) (
\hat{A}_{-}\hat{S} _{+}+\hat{A}_{+}\hat{S}_{-} )$. However, in
general, the term $\hat{V}_{D} = \hat{A}_{z}\hat{S}_{z}$ cannot be
neglected. Even on resonance, where $ B_{0}  = - (1/g^{\ast }\mu
_{B}) \langle \hat{A} _{z} \rangle$, one has $\langle
\hat{A}_{z}^{2} \rangle - \langle \hat{A} _{z} \rangle ^{2} \neq 0$.
In conclusion, the Overhauser field felt by the electron spin cannot
be fully compensated. To include this effect we rewrite the
$zz$-part of the interaction as $\hat A_z = \bar\alpha \hat{J}_{z} +
\delta \hat A_z =\bar \alpha \sum_{i=1}^{N}\hat{I}_z^{(i)}+ \sum_i
(\alpha_i - \bar \alpha) \hat{I}_z^{(i)}$, so that the on-resonance
Hamiltonian becomes
\begin{equation}
\widehat{\widetilde{H}}_{\mathrm{QD}}=\frac{1}{2}\left(
\hat{A}_{-}\hat{S}_{+}+\hat{A}_{+} \hat{S}_{-}\right) +\bar \alpha
\hat{S}_{z}(\hat{J}_{z} - \langle \hat{J}_{z} \rangle_0  ) +
\hat{V}_{zz}. \label{ham2}
\end{equation}
Here $\langle \hat{J}_{z} \rangle_0 $ is the expectation value
with respect to the fully polarized state, $\bar \alpha = \sum_i
\alpha_i/N$ is the average coupling constant (which equals $A/N$
due to the normalization of the electron wave function), and
$\hat{V}_{zz} = \hat{S}_z \delta \hat A_z$. The latter term, i.e.
the inhomogeneous contribution to the $z$-term of the interaction,
can be treated perturbatively. As we show later, the homogeneous
$zz$-term $\hat{S}_{z}\hat{J}_{z}$ is neglected with respect to
the flip-flop term due to a small factor $1/\sqrt{N}$. The term
$\hat{V}_{zz}$ is even smaller and negligible, because it depends
additionally on an inhomogeneity parameter, like the variance of
the (smooth) electron wave function.

In this work we consider $I=1/2$ nuclear spins in a spherical
quantum dot, and
\begin{equation}
\left| \psi ( \vec{r}) \right|^2 \propto e^{-\frac{r^2}{r_0^2}},
\end{equation}
as the electron envelope function~\cite{Schliemann} of size $r_0$.

\subsection{Electron and Nuclear Spin Dynamics\label{enucd}}

The inhomogeneous coupling of the electron spin to the nuclear spins
does not allow for a general analytical solution of the dynamics.
The assumption of a fully polarized initial nuclear spin state
reduces the dimension of the relevant Hilbert space from $2^{N+1}$
to $N+1$ and therefore the problem is readily analyzed. On the other
hand, a more general initial state of nuclear spins vastly increases
the dimension of the Hilbert space that the system will visit along
the evolution. Motivated by the proposals for quantum information
storage via the hyperfine interaction~\cite{Taylor}, we will study
situations of imperfect, but high, polarization. We analyze the case
where the dynamics involves one excitation in the nuclear spin
system, which we will refer to as ``defect''. We show, with methods
similar to the ones developed in Sec.~II, that the relevant Hilbert
space remains still small.

The most general pure state describing this situation can be
written as
\begin{equation}
\left\vert \mathbf{1}\right\rangle _{n}=\sum_{j=1}^{N}a_{j}\left\vert
\mathbf{1}_{j}\right\rangle _{n},  \label{unodefecto}
\end{equation}
with $\sum_{j=1}^{N} |a_j|^2 =1$, and $\left\vert \mathbf{1}
_{j}\right\rangle _{n}$ represents the state $\left\vert
\downarrow \downarrow ...\uparrow _{j}...\downarrow \downarrow
\right\rangle _{n}$ with an inverted nuclear spin at position
$r_{j}$. In particular we consider the case of a uniformly
distributed defect $a_{j}=1/\sqrt{ N}$ and a localized defect at
position $j_{0}$,
\begin{equation} a_{j}=C\left( \Gamma /2\right)
^{2}/\left[ \left( j-j_{0}\right) ^{2}+\left( \Gamma /2\right)
^{2}\right],\label{locateddefect}
\end{equation} where $C$ is a normalization constant
and $\Gamma $ characterizes the width of the distribution.

Let us consider two possible initial states for the electron and
the nuclear spins with a single imperfection,
\begin{eqnarray}
\left\vert \downarrow \right\rangle _{e}\left\vert \mathbf{1}\right\rangle
_{n} &\equiv &\left\vert \downarrow \mathbf{1}\right\rangle , \label{cond1} \\
\left\vert \uparrow \right\rangle _{e}\left\vert
\mathbf{1}\right\rangle _{n} &\equiv &\left\vert \uparrow
\mathbf{1}\right\rangle . \label{cond2}
\end{eqnarray}
In order to solve the dynamical evolution of the composite system,
we employ the recipe of following the relevant Hilbert space, as
described in Sec.~II. We observe that, under the action of the
Hamiltonian of Eq.~(\ref{ham2}), the initial state $\left\vert
\downarrow \mathbf{1}\right\rangle $ is coupled to the state
$\left\vert \uparrow \mathbf{0}\right\rangle$ via
$\hat{S}_{+}\hat{A}_{-}\left\vert \downarrow
\mathbf{1}\right\rangle =\gamma \left\vert \uparrow
\mathbf{0}\right\rangle$, with $\left\vert \mathbf{0}\right\rangle
\equiv \left\vert \mathbf{0}\right\rangle_{n} = \left\vert
\downarrow \downarrow ...\downarrow \downarrow \right\rangle _{n}$
and
\begin{equation}
\gamma =\sum_{i}a_{i}\alpha _{i}.\label{gamma}
\end{equation}
The state $\left\vert \uparrow \mathbf{0}\right\rangle$ will be
coupled through the term $\hat{S}_{-}\hat{A}_{+}$ to a state with
one excitation, but different from $\left\vert \downarrow
\mathbf{1} \right\rangle$. Similar to the formalism developed in
Sec.~II, we can write $\hat{S}_{-}\hat{A}_{+}\left\vert \uparrow
\mathbf{0}\right\rangle =\gamma \left\vert \downarrow
\mathbf{1}\right\rangle +\beta \left\vert \downarrow
\mathbf{1}_{\bot }\right\rangle$, where state.
\begin{equation}
\left\vert \mathbf{1}_{\bot }\right\rangle =\frac{1}{\beta }\left(
\sum_{i}\alpha _{i}\left\vert \mathbf{1}_{i}\right\rangle -\gamma \left\vert
\mathbf{1}\right\rangle \right)
\end{equation}
is orthogonal to state $\left\vert \mathbf{1}\right\rangle$, with
$\beta = \sqrt{\left( N_{\mathbf{0}}\right) ^{2}-\gamma ^{2}}$ and
$N_{\mathbf{0}} = \sqrt{\sum\limits_{i}\alpha _{i}^{2}}$. Finally,
we observe that $\hat{S}_{+}\hat{A}_{-}\left\vert \downarrow
\mathbf{1}_{\perp }\right\rangle =\beta \left\vert \uparrow
\mathbf{0}\right\rangle$. In summary, the exchange terms yield
\begin{eqnarray}
\hat{S}_{+}\hat{A}_{-}\left\vert \downarrow \mathbf{1}\right\rangle
&=&\gamma \left\vert \uparrow \mathbf{0}\right\rangle , \nonumber \\
\hat{S}_{-}\hat{A}_{+}\left\vert \uparrow \mathbf{0}\right\rangle
& = & \gamma \left\vert \downarrow \mathbf{1}\right\rangle +\beta
\left\vert \downarrow
\mathbf{1}_{\bot }\right\rangle , \nonumber\\
\hat{S}_{+}\hat{A}_{-}\left\vert \downarrow \mathbf{1}_{\bot
}\right\rangle &=&\beta \left\vert \uparrow
\mathbf{0}\right\rangle .
\end{eqnarray}
Furthermore, the $zz$-interaction yields matrix elements
\begin{eqnarray}
\left\langle \downarrow \mathbf{1}\right\vert
\bar{\alpha}\hat{S}_{z}(\hat{J}_{z} - \langle \hat{J}_{z}
\rangle_0  )\left\vert
\downarrow \mathbf{1}\right\rangle &=&-\frac{\bar{\alpha}}{2} , \nonumber \\
\left\langle \uparrow \mathbf{0}\right\vert
\bar{\alpha}\hat{S}_{z}(\hat{J}_{z} - \langle \hat{J}_{z}
\rangle_0  )\left\vert \uparrow
\mathbf{0}\right\rangle & = & 0 , \nonumber\\
\left\langle \downarrow \mathbf{1}_{\bot }\right\vert
\bar{\alpha}\hat{S}_{z}(\hat{J}_{z} - \langle \hat{J}_{z}
\rangle_0  )\left\vert \downarrow \mathbf{1}_{\bot }\right\rangle
&=&-\frac{\bar{\alpha}}{2} .
\end{eqnarray}
Therefore, for the initial condition of Eq.~(\ref{cond1}) the
system evolves only in the subspace spanned by $\left\{ \left\vert
\downarrow \mathbf{1} \right\rangle ,\left\vert \uparrow
\mathbf{0}\right\rangle ,\left\vert \downarrow \mathbf{1}_{\bot
}\right\rangle \right\} $. The temporal evolution in this subspace
can be exactly solved, obtaining $\left\vert \Psi _{\downarrow
}\left( t\right) \right\rangle _{e-n}=a_{1}(t)\left\vert
\downarrow \mathbf{1}\right\rangle +b_{1}(t)\left\vert \uparrow
\mathbf{0} \right\rangle +c_{1}(t)\left\vert \downarrow
\mathbf{1}_{\bot }\right\rangle $, where the probability
amplitudes are given by
\begin{eqnarray}
a_{1}(t) & = & \!\! 1 + \frac{\gamma ^{2}\exp \left(
i\bar{\alpha}t/4\right) }{\Omega
^{2}\delta }[\delta \cos \left( \delta t/2\right) \nonumber \\
&&+i\frac{\bar{\alpha}}{2}\sin \left( \delta t/2\right) -\delta \exp \left(
-i\bar{\alpha}t/4\right) ]  , \nonumber\\
b_{1}(t) & = & \!\! -i\frac{\gamma }{\delta }\exp \left(
i\bar{\alpha}t/4\right)
\sin \left( \delta t/2\right) , \nonumber\\
c_{1}(t) & = & \!\! \frac{\gamma \beta \exp \left(
i\bar{\alpha}t/4\right) }{\Omega ^{2}\delta }[\delta \cos \left(
\delta t/2\right) +i\frac{\bar{\alpha}}{2}
\sin \left( \delta t/2\right)  \nonumber \\
&&-\delta \exp \left( -i\bar{\alpha}t/4\right) ] .
\end{eqnarray}
Here, $\Omega =\sqrt{\gamma ^{2}+\beta ^{2}}=N_{0}$ and $\delta
=\sqrt{\left( \bar{\alpha}/2\right) ^{2} + \Omega ^{2}}$ plays the
role of a generalized Rabi frequency for a process with detuning
$\bar \alpha$. The $zz$-interaction thus leads to a small detuning
for the exchange of excitation between electron and nuclear spins:
We have that $\Omega = \sqrt{\gamma^2 + \beta^2}=N_0 =
\sqrt{\sum_i \alpha_i^2}$. The $\alpha_i$ are $\mathcal{O}(1/N)$,
such that $\sum_i \alpha_i^2 = \mathcal{O}(1/N)$. When $N \gg 1$,
then $1/N=\bar{\alpha} \ll \Omega=\mathcal{O}(1/\sqrt{N})$. It is
important to notice that in the case of an initially fully
polarized nuclear state $\left\vert \mathbf{0}\right\rangle $ and
the electron spin in $ \left\vert \downarrow \right\rangle _{e}$
the system is not affected by the exchange interaction terms
$\hat{S}_{-}\hat{A}_{+}$ and $\hat{S}_{+}\hat{A}_{-}$.

\begin{figure}
\hspace*{-0.2cm}\includegraphics[width=80mm,height=60mm]{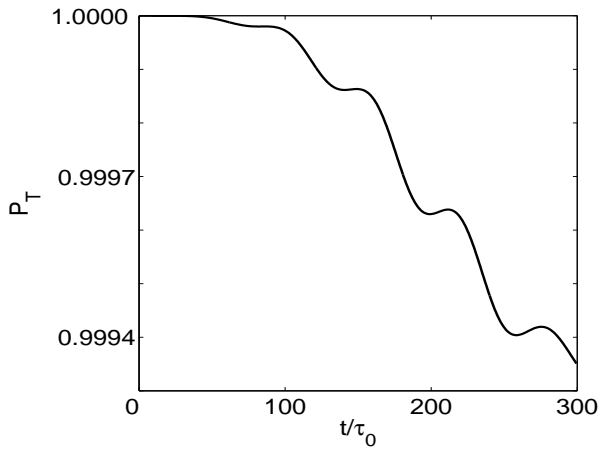}\vspace*{0.25cm}
\caption{Total population $P_{\rm T}$ of the states $\left\{ \left\vert
\uparrow \mathbf{1} \right\rangle ,\left\vert \downarrow
\mathbf{2}\right\rangle ,\left\vert \uparrow
\mathbf{1}_{p}\right\rangle ,\left\vert \downarrow \mathbf{2}
_{p}\right\rangle \right\} $, with $N=10^{3}$ and defect
distribution $ a_{j}=1/\protect\sqrt{N}$.} \label{pob4statesN1000}
\end{figure}

On the other hand, if the system is initially in the state
$\left\vert \uparrow \right\rangle _{e}\left\vert
\mathbf{1}\right\rangle _{n}$, the coupling through the Hamiltonian
of Eq.~(\ref{ham2}) does not lead to a closed Hilbert space. Then,
some approximations are necessary in order to obtain solutions for
the evolution of the overall system dynamics. Starting from
$\left\vert \uparrow \mathbf{1}\right\rangle$, we have
\begin{equation}
\hat{S}_{-}\hat{A}_{+}\left\vert \uparrow \mathbf{1}\right\rangle
=\left\vert \downarrow \right\rangle \sum_{i<j}b_{ij}\left\vert
\mathbf{2} _{ij}\right\rangle , \label{op21}
\end{equation}
where $b_{ij}=a_{i}\alpha _{j}+a_{j}\alpha _{i}$ and $\left\vert
\mathbf{2} _{ij}\right\rangle =\left\vert \downarrow \downarrow
...\uparrow _{i}...\uparrow _{j}...\downarrow \downarrow
\right\rangle $. Defining
\begin{equation}
\left\vert \mathbf{2}\right\rangle =\sum_{i<j}c_{ij}\left\vert
\mathbf{2} _{ij}\right\rangle ,
\end{equation}
with $c_{ij}=b_{ij}/N_{\mathbf{2}}$ and $N_{\mathbf{2}}=\sqrt{
\sum\limits_{i<j}b_{ij}^{2}}$, we observe that the state
$\left\vert \downarrow \mathbf{2}\right\rangle $ is coupled only
through
\begin{equation}
\hat{S}_{+}\hat{A}_{-}\left\vert \downarrow
\mathbf{2}\right\rangle =\sum_{i}(\sum_{j\neq i}c_{ij}\alpha
_{j})\left\vert \uparrow \mathbf{1} _{i}\right\rangle \equiv
\left\vert \uparrow \phi _{1}\right\rangle .
\end{equation}
Here, $\left\vert \phi _{1}\right\rangle $ can be seen as having
one component along the state $\left\vert \mathbf{1}\right\rangle
$ and other component along orthogonal state $\left\vert
\mathbf{1}_{p}\right\rangle $ , such that
\begin{equation}
\label{phi} \left\vert \phi _{1}\right\rangle =\left\langle
\mathbf{1}\right. \left\vert \phi _{1}\right\rangle \left\vert
\mathbf{1}\right\rangle +\sqrt{ \left\langle \phi _{1}\right.
\left\vert \phi _{1}\right\rangle -\left\langle \mathbf{1}\right.
\left\vert \phi _{1}\right\rangle ^{2}} \left\vert
\mathbf{1}_{p}\right\rangle ,
\end{equation}
where $\left\vert \mathbf{1}_{p}\right\rangle =\left( 1/N_{
\mathbf{1}_{p}}\right) \left( \left\vert \phi _{1}\right\rangle
-\left\langle \mathbf{1}\right. \left\vert \phi _{1}\right\rangle
\left\vert \mathbf{1}\right\rangle \right) $, with
$N_{\mathbf{1}_{P}}=\sqrt{ \left\langle \phi _{1}\right.
\left\vert \phi _{1}\right\rangle -\left\langle \mathbf{1}\right.
\left\vert \phi _{1}\right\rangle ^{2}}$. Following this
procedure, we have
\begin{equation}
\hat{S}_{-}\hat{A}_{+}\left\vert \uparrow
\mathbf{1}_{p}\right\rangle =\left\vert \downarrow
\mathbf{2}\right\rangle +N_{\mathbf{2}_{p}}\left\vert \downarrow
\mathbf{2}_{p}\right\rangle \equiv \left\vert \uparrow \phi
_{2}\right\rangle ,
\end{equation}
where $\left\vert \mathbf{2}_{p}\right\rangle =\left( 1/N_{
\mathbf{2}_{P}}\right) \left( \left\vert \phi _{2}\right\rangle
-\left\langle \mathbf{2}\right. \left\vert \phi _{2}\right\rangle
\left\vert \mathbf{2}\right\rangle \right) $, with $\left\vert
\phi _{2}\right\rangle = \hat{A}_{+}\left\vert
\mathbf{1}_{p}\right\rangle $ and $N_{\mathbf{2}_{P}}=
\sqrt{\left\langle \phi _{2}\right. \left\vert \phi
_{2}\right\rangle -\left\langle \mathbf{2}\right. \left\vert \phi
_{2}\right\rangle ^{2}}$. In order to implement a semi-analytical
description for the overall system dynamics we have to truncate
the Hilbert space. This can be accomplished by approximating the
action of $\hat{S}_{+}\hat{A}_{-}$\ over $\left\vert
\mathbf{2}_{p}\right\rangle $ such that
\begin{eqnarray}
\hat{S}_{+}\hat{A}_{-}\left\vert \downarrow
\mathbf{2}_{p}\right\rangle &=&N_{\mathbf{2}_{P}}\left\vert
\uparrow \mathbf{1}_{p}\right\rangle + N_{\mathbf{1}pp}\left\vert
\uparrow \mathbf{1}_{pp}\right\rangle  \notag \\
&\simeq &N_{\mathbf{2}_{P}}\left\vert \uparrow \mathbf{1}_{p}
\right\rangle, \label{aprox1}
\end{eqnarray}
where the state $\left\vert \uparrow \mathbf{1}_{pp}\right\rangle
$ is similarly defined in terms of $\left\vert \uparrow \mathbf{1}
_{p}\right\rangle $. This approximation can be justified from
Fig.~\ref{pob4statesN1000} where we have plotted the total
occupation probability of states within the subspace of first
orthogonal states $\left\vert \uparrow \mathbf{1}\right\rangle
,\left\vert \downarrow \mathbf{2}\right\rangle ,\left\vert
\uparrow \mathbf{1} _{p}\right\rangle ,\left\vert \downarrow
\mathbf{2}_{p}\right\rangle $. As all of them are eigenstates of
the $\hat{J}_z$ term, its effect is again a small detuning. These
results have been obtained by numerical calculation considering a
12-dimensional truncated Hilbert space up to $\left\vert
\downarrow \mathbf{2}_{ppppp}\right\rangle $. However, if we use a
Hilbert space truncated up to $\left\vert \downarrow \mathbf{2}
_{pp}\right\rangle $, the total occupation probability of the
mentioned subspace is practically the same as shown in
Fig.~\ref{pob4statesN1000}.

We have seen before that the typical timescale for the exchange of
an excitation between the electron and nuclei is given by the
inverse of the generalized Rabi frequency
\begin{equation}
\label{transfertime}
\tau := \pi /\Omega=\pi/N_0.
\end{equation}
From Fig.~\ref{pob4statesN1000} we deduce that the dynamics of the
composite system is well described for times beyond $\tau$, which is
important for our subsequent analysis of the quantum memory. In
particular, for the case $N=10^{3}$, this time is approximately $\tau \approx 68 \tau_0$, with
$\tau_0 = \hbar / A$, and we are allowed to restrict the
dynamics to the subspace spanned by $\left\{ \left\vert \uparrow
\mathbf{1}\right\rangle ,\left\vert \downarrow
\mathbf{2}\right\rangle ,\left\vert \uparrow \mathbf{1
}_{p}\right\rangle ,\left\vert \downarrow
\mathbf{2}_{p}\right\rangle \right\} $. This is an important feature
of our method, since numerical calculations are significantly
simplified.

\subsection{Quantum Information Storage}

\begin{figure}[t]
\includegraphics[width=65mm,height=60mm]{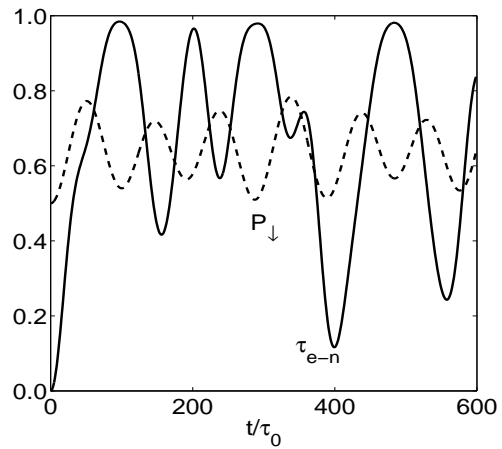}\vspace*{0.0cm}
\caption{Evolution of the tangle $\protect\tau
_{e-n}$ (solid line) and \ population $P_{\downarrow }$ of
$\left\vert \downarrow \right\rangle _{e}$ (dashed line) for a
uniformly distributed defect. The initial state is $\left\vert
\Psi \right\rangle =1/\protect\sqrt{2}\left( \left\vert \uparrow
\right\rangle +\left\vert \downarrow \right\rangle \right)
_{e}\otimes \left\vert \mathbf{1 }\right\rangle _{n}$, $N=10^{3}$,
and $\tau \sim 68 \tau_0$.}
\label{tangle4states600N1000}
\end{figure}

In Refs.~\cite{Taylor} a long-lived quantum memory based on the
capability that an electronic spin state can be reversibly mapped
into a fully-polarized nuclei state was proposed. This dynamics can
be expressed as
\begin{equation}
\left( u\left\vert \uparrow \right\rangle + v \left\vert
\downarrow \right\rangle \right) \otimes \left\vert \mathbf{0}
\right\rangle \rightarrow \left\vert \downarrow \right\rangle
\left( v \left\vert \mathbf{1} \right\rangle +iu\left\vert
\mathbf{0}\right\rangle \right)   \label{lukin}
\end{equation}
where $\left\vert \mathbf{1}\right\rangle =\left( 1/N_{\mathbf{0}
}\right) \sum_{i}\alpha _{i}\left\vert \mathbf{1}_{i}\right\rangle
$. This coherent mapping is effected by pulsing the applied field
from $g^{\ast }\mu _{B}B_{\mathrm{eff}} \gg \hat{V}_{\Omega }$ to
$B_{ \mathrm{eff}}\sim 0$ for a time $\tau =\pi /N_{\mathbf{0}}$.
Here, we study this dynamics in the context of an initial
distributed single defect in the nuclear spin state. For the
transfer process to be successful it is necessary that the
electronic state factorizes from the nuclear spin state, thus
requiring that at some point of the evolution
\begin{eqnarray}
\left( u\left\vert \uparrow \right\rangle + v\left\vert \downarrow
\right\rangle \right) \otimes \left\vert \mathbf{1}\right\rangle
_{n}\rightarrow \left\{
\begin{array}{c}
\left\vert \downarrow \right\rangle _{e}\otimes \left\vert
\psi^{\downarrow}
_{u,v}\right\rangle _{n} \\
\left\vert \uparrow \right\rangle _{e}\otimes \left\vert
\psi^{\uparrow} _{u,v}\right\rangle _{n} \, .
\end{array}
\right. \label{mem2}
\end{eqnarray}
In order to verify if this is possible we analyze the populations
of the electron spin and study the entanglement between the
electron and the nuclear spins. Since the overall system is in a
pure state we can use as an entanglement measure the
tangle~\cite{Rungta}, $ \tau _{AB}$=$2\nu _{A}\nu _{B}\left[
1-\text{\textrm{tr}}\left( \rho _{A}^{2}\right) \right] $, where
$\rho _{A}$ is the reduced density operator after tracing over
subsystem $B$, and $\nu _{A}$ and $\nu _{B}$ are arbitrary scale
factors set to $\nu _{A}=\nu _{B}=1$.

We first consider the case of a uniformly distributed initial
nuclear spin excitation. In Fig.~\ref{tangle4states600N1000} we show
the electron spin population $p_\downarrow$ and the tangle between
the electron and $N=10^{3}$ nuclear spins. We observe that, for the
case of a uniformly distributed defect, there is no time for which
the composite state evolves to a factorized state as in
Eq.~(\ref{mem2}). The electron spin population is never maximal or
minimal (in both cases the tangle would vanish), meaning that there
is no high-fidelity information transfer to the nuclei.

\begin{figure}[t]
\includegraphics[width=70mm,height=65mm]{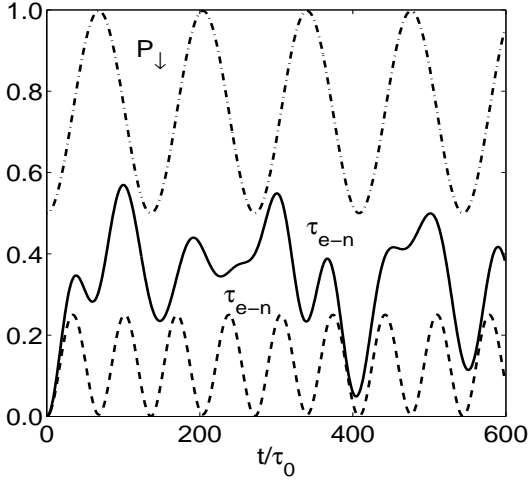}\vspace*{0.0cm}
\caption{Evolution of the tangle $\protect\tau
_{e-n}$ for a defect distribution peaked at the edge of the
quantum dot. Shown are the electron spin population $P_{\downarrow
}$ (dot-dashed line) and the tangle (dashed line) for a defect
distribution width $ \Gamma =N/50$, and the tangle for the width
$\Gamma =N$ (solid line) for $N=10^3$ and the initial state
$\left\vert \Psi \right\rangle =1/\protect\sqrt{2}\left(
\left\vert \uparrow \right\rangle +\left\vert \downarrow
\right\rangle \right) _{e}\otimes \left\vert \mathbf{1}
\right\rangle _{n}$.} \label{distcenN}
\end{figure}

We study now the transfer process for the case of a localized defect
as shown in Eq.~(\ref{locateddefect}). When the defect is peaked in
the center of the dot, information storage is seriously hindered.
However, if the defect is more probably located at an extreme of the
quantum dot, the coupling of the defect-spin to the electron is
weak, and separability is reached even for rather broad defect
distributions, as seen in Fig.~\ref{distcenN}. The necessary
conditions for a successful transfer are fulfilled after a time $t
\approx 68 \tau_0$, which is very close to the transfer time
for an initial fully polarized nuclear spin state given by
Eq.~(\ref{transfertime}).

\begin{figure}
\hspace*{-0.3cm}\includegraphics[width=70mm,height=65mm]{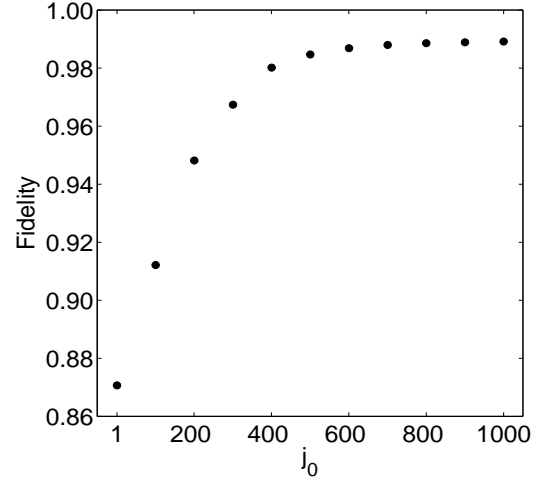}\vspace*{0.0cm}
\caption{Fidelity of the quantum memory for $N=10^{3}$ nuclear spins
and different $j_{0}$, $1 \leq j_{0} \leq N$.} \label{figfidelity}
\end{figure}

\vspace*{0.9cm}

\begin{figure}
\vspace*{0.45cm}
\hspace*{-0.4cm}\includegraphics[width=70mm,height=65mm]{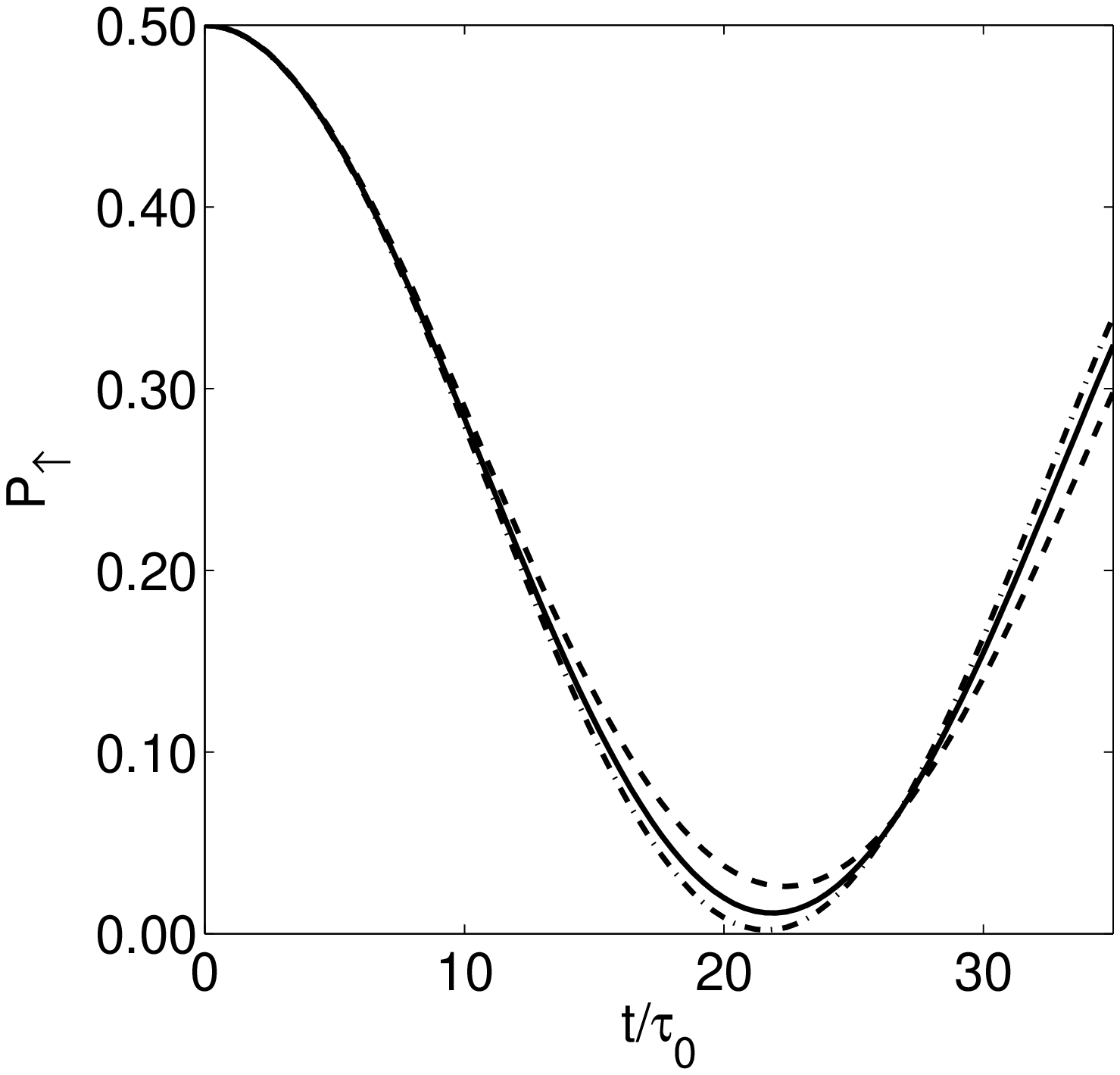}\vspace*{0.0cm}
\caption{Evolution of the population $P_{\uparrow }$ of the
electronic spin state $\left\vert \uparrow \right\rangle _{e}$ for
$N=40$. The initial state for the electron spin is $\left\vert
\Psi \right\rangle _{e}=1/\protect\sqrt{ 2}\left( \left\vert
\uparrow \right\rangle +\left\vert \downarrow \right\rangle
\right) _{e}$ and $\protect\rho _{N}=\sum_{j=1}^{N}a_{j}
\protect\hat{\sigma} _{+}^{\left( j\right) }\left\vert
\mathbf{0}\right\rangle \left\langle \mathbf{0}\right\vert
\protect\hat{\sigma} _{-}^{\left( j\right) }$ for the nuclear
spins. Solid line: $a_{j}=1/N$; Dashed line: $a_{j}=[C\left(
\Gamma /2\right) ^{2}/(\left( j-j_{0}\right) ^{2}+\left( \Gamma
/2\right) ^{2})]^{2}$, with $\Gamma =N$ \, $j_{0}=1$, and  $C$ a
normalization constant. Dot-dashed line: same as for dashed line
but $j_{0}=N$. } \label{figrhon}
\end{figure}

The performance of the quantum information storage protocol is
analyzed now in a more quantitative manner by simulating a complete
cycle of write-in, storage and retrieval. In the first step the
electron spin in state $|\psi_i\rangle$ interacts with the nuclei
and after a time $\tau$ the evolution is interrupted and the
electron traced out. Then a fresh electron in state
$|\downarrow\rangle$ interacts with the nuclei and after another
time $\tau$ the nuclei are traced out. The fidelity $F=\langle
\psi_i | \rho_f | \psi_i\rangle$ between this final electron spin
state $\rho_f$ and the initial state is plotted in
Fig.~\ref{figfidelity} for various defect distributions, where $F$
has been calculated by averaging on the electron spin Bloch sphere.

In Fig.~\ref{figrhon} we consider the nuclear spins in a mixed state
where electron spin populations are displayed. This figure shows
that the electron spin performs very regular Rabi oscillations that
deviate only slightly from the ideal situation with no defect
present. In particular the performance is better than in the
corresponding cases of coherently distributed defects. This is to be
expected as the quantum information storage is sensitive to the
coherences of the nuclear spins state, and they are absent in this
mixed state case. Note that in Fig.~\ref{figrhon} the initial number
of nuclear spin excitations is fixed, at variance with a thermal
distribution. In this case the visited Hilbert space scales as $N^2$
and not as $2^{N+1}$, allowing an exact calculation.

\begin{figure}
\hspace*{-0.4cm}\includegraphics[width=70mm,height=65mm]{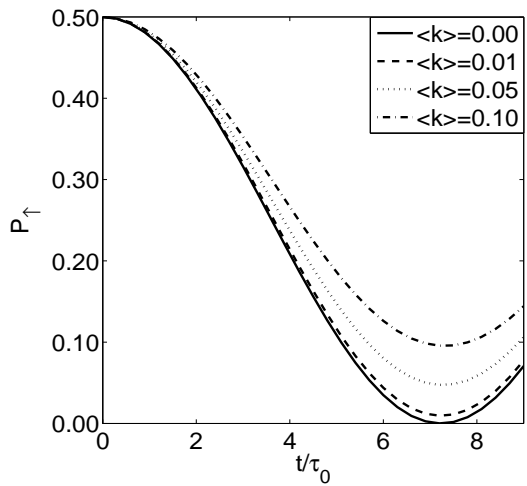}\vspace*{0.0cm}
\caption{Evolution of population $P_{\uparrow }$ of electronic
spin state $ \left\vert \uparrow \right\rangle _{e}$ for an initial
thermal distribution for each nuclear spin $N=10$ and different
$\left\langle k\right\rangle $. The initial electron spin state is
$\left\vert \Psi \right\rangle =1/ \protect\sqrt{2}\left( \left\vert
\uparrow \right\rangle +\left\vert \downarrow \right\rangle \right)
_{e}$. For this case, $\left\langle k\right\rangle =\exp \left(
-\protect\beta \hbar \protect\omega \right) /\left( 1+\exp \left(
-\protect\beta \hbar \protect\omega \right) \right) .$}
\label{thermalsingle}
\end{figure}

Another possible physical scenario is that of an initial nuclear
system at finite temperature. Our method is not directly applicable
to that situation because already the initial state covers a large
part of the Hilbert space. Nevertheless, for the sake of
completeness, we include a brief analysis of this case in the
context of a quantum memory in the presence of defects. Consider a
thermal state for each nuclear spin
\begin{eqnarray}
\rho_{\text{\textrm{th}}}^{i}=\frac{e^{-\hbar \omega \hat{\sigma}
_{z}^{i}\beta }}{\text{\textrm{tr}}(e^{-\hbar \omega
\hat{\sigma}_{z}^{i}\beta })}  \label{ter2}
\end{eqnarray}
with $\beta =1/k_{B}T$, where $T$ is the associated temperature. The
average population of each spin is $\langle k \rangle = \langle
\hat{\sigma}_z^{(i)} \rangle + 1/2$. As for the initial mixed state
case of Fig.~8, it is not possible to define collective states as
was the case for pure initial states. The results of our direct
simulations taking into account the whole Hilbert space are shown in
Fig.~\ref{thermalsingle}, where we show the population of the
electronic spin state $\left\vert \uparrow \right\rangle$. For
nuclear polarizations above 95\% the contrast of the Rabi
oscillations of the electron spin deviate from the ideal situation
by only a few percent. The scaling of the error with the particle
number $N$ is certainly beyond the scope of this article.

\section{CONCLUSIONS}

We have developed a technique, based on suitable inspection and
truncation of the associated Hilbert space, that allows to follow
the quantum evolution of interacting systems with inhomogeneous
coupling in an accurate and controlled manner. As a first general
case, we illustrated the proposed method for the inhomogeneous
Tavis-Cummings model and showed how the statistical properties may
change appreciably depending on the spatial coupling distribution.
As a second major example, we studied the dynamics of an electron
spin interacting inhomogeneously with a system of polarized nuclear
spins in a quantum dot, previously considered as a quantum memory
candidate. We found that the reliability of the storage process
depends strongly on the presence and position of a single
distributed defect in the polarized nuclei state. The proposed
technique may be easily extended to other inhomogeneously
interacting systems and may prove to be accurate and efficient when
interaction times are known beforehand, as is the case of pulses.

\begin{acknowledgments}
The authors acknowledge useful discussions with G. Giedke. C.E.L. is
financially supported by MECESUP USA0108, H.C. by DFG SFB 631,
J.C.R. by Fondecyt 1030189 and Milenio ICM P02-049F, and E.S. by DFG SFB 631 and EU RESQ and EuroSQIP projects.
\end{acknowledgments}

\end{document}